\documentclass[pra,showpacs,superscriptaddress,twocolumn]{revtex4}

\usepackage{amsmath}
\usepackage{amsthm}
\usepackage{amssymb}
\usepackage{amsfonts}
\usepackage{graphicx}

\DeclareMathOperator{\Tr}{Tr}

\newcommand{\ket}[1]{|{#1}\rangle}
\newcommand{\kb}[1]{|{#1}\rangle \langle{#1}|}

\newcommand{\kbt}[2]{|{#1}\rangle \langle{#2}|}

\makeatletter

\newtheorem{thm}{Theorem}

\begin{document}
\title{Emergence of Randomness and Arrow of Time in Quantum Walks}
\author{Yutaka Shikano}
\email{shikano@mit.edu}
\affiliation{Department of Physics, Tokyo Institute of Technology, Meguro, Tokyo 152-8551, Japan}
\affiliation{Department of Mechanical Engineering, Massachusetts Institute of Technology, 
Cambridge, MA 02139, USA}
\author{Kota Chisaki}
\email{d07sb103@ynu.ac.jp}
\affiliation{Department of Applied Mathematics, Faculty of Engineering, Yokohama National University, 
Hodogaya, Yokohama 240-8501, Japan}
\author{Etsuo Segawa}
\email{segawa.e.aa@m.titech.ac.jp}
\affiliation{Department of Value and Decision Science, Tokyo Institute of Technology, Meguro, Tokyo 152-8551, Japan}
\author{Norio Konno}
\email{konno@ynu.ac.jp}
\affiliation{Department of Applied Mathematics, Faculty of Engineering, Yokohama National University, 
Hodogaya, Yokohama 240-8501, Japan}
\date{\today}
\begin{abstract}
Quantum walks are powerful tools not only to construct the quantum speedup algorithms but also to describe specific models in physical processes. 
Furthermore, the discrete time quantum walk has been experimentally realized in various setups. We apply the concept 
of the quantum walk to the problems in quantum foundations. We show that randomness and the arrow of time in the quantum walk gradually 
emerge by periodic projective measurements from the mathematically obtained limit distribution under the time scale transformation.
\end{abstract}
\maketitle 
\section{Introduction}
The discrete time quantum walk (DTQW), which is a quantum analogue of the random walk (RW)~\cite{Aharonov,comment}, has been expected to be a powerful tool 
in various fields, especially quantum computation~\cite{Childs2,Lovett,VA}, physics~\cite{Chandrashekar, Oka}, and mathematics~\cite{Kempe, KonnoRev},  
and has been experimentally realized~\cite{Ryan,Karski,Photon,Roos}. 
The important properties are an inverted-bell shaped limit distribution, which is extremely different from the normal distribution obtained 
in the RW, and the faster diffusion process than the RW because of a coherent superposition 
and a quantum interference~\cite{KonnoQL0, KonnoQL1}. In this paper, we derive the limit distribution of the DTQW with 
the periodic position measurement (PPM) under the time scale transformation. We show that randomness and the arrow 
of time gradually emerge under the time scale transformation from the mathematically obtained limit distribution. 
The result means that randomness in QWs is quantified and gives us a new insight into the interpretation of the projective measurement.

Our addressed issues on randomness in QWs and the arrow of time have the following background. 
First, it is well known that randomness results from environment in the RW. 
Although some researchers often call the QW a quantum ``random" walk due to its original definition~\cite{Aharonov,comment}, 
this terminology is abuse of words since the QW is not probabilistic but 
deterministic due to the time evolution for the closed quantum system. As far as we know, no one has yet analyzed a quantification of randomness in QWs. 
Environment, of which the description corresponds to that of quantum measurement in quantum mechanics~\cite{NC}, seems intuitively to be the origin 
of randomness. In such studies, the description of the periodic quantum measurement is used 
and the quantum-classical transition, which means an immediate change from the QW to the RW due to 
the decoherence effect, is shown~\cite{Brun, Kendon, Kendon2, Zhang}. However, the contribution to the quantum walk 
behavior from environment has not been shown analytically.

Second, we apply the concept of the QW to quantum foundations, especially the problem of quantum measurement. 
This problem has been discussed for a long time. Its central question is what is the physical origin of the time asymmetry in quantum 
measurement~\cite{time}. According to the ``Copenhagen interpretation", the projective measurement produces 
the arrow of time since its description is time asymmetric~\cite{comment3}. Furthermore, it seems to be natural to consider that the description of 
the quantum dynamics with many projective measurements can be taken as the Markov process since the projective measurement is to forget 
the measurement outcome~\cite{Open}. In other words, the arrow of time is uniformly produced by the projective measurements for 
any time scale. While our proposal is not complete solution on the arrow of time, this 
gives us a new insight from the DTQW. In the paper, we discuss the above problems with 
the mathematical model of the DTQW.

The paper is organized as follows. In Sec.~\ref{review}, we recapitulate the definition of the DTQW.
In Sec.~\ref{ours}, we define a specific model of the DTQW with periodic position measurement. We show the mathematical result on the limit distribution and give 
an interpretation to our mathematical result in the randomness and the arrow of time. Section~\ref{conc} is devoted to the summary.
\section{Discrete Time Quantum Walk} \label{review}
Let us mathematically define the one-dimensional DTQW~\cite{Ambainis} as follows. 
First, we prepare the position and the coin states denoted as $\kb{x}$ and $\rho_0$, respectively, corresponding to the 
quantization of the RW~\cite{Aharonov}. 
Here, we assume that the position is the one-dimensional discretized lattice denoted as $\mathbb{Z}$ and the coin state is 
a qubit with the orthonormal basis, $\ket{L} = ( 1, 0 )^{T}$ and $\ket{R} = ( 0, 1 )^{T}$, where $T$ is transposition. 
To simplify the discussion, we assume that the initial state is  
localized at the origin ($x=0$) with the mixed state $\rho_0 = ( \kb{L} + \kb{R})/2$ as the coin state throughout this paper. Second, the time evolution of 
the QW is described by a unitary operator $U$. A quantum coin flip corresponding to the coin flip in the RW is described by a unitary 
operator $H \in U(2)$ acting on the coin state given by 
\begin{align}
H & := a \kbt{L}{L} + b \kbt{L}{R} + c \kbt{R}{L} + d \kbt{R}{R} \notag \\ 
& = \left( \begin{array}{cc} a & b \\ c & d \end{array} \right)
\end{align}
with $|a|^2 + |c|^2 = 1$, $a \bar{b} + c \bar{d} = 0$, $c = - \Delta \bar{b}$, 
$d = \Delta \bar{a}$, and $|\Delta| := | \det U| = 1$,
noting that $abcd \neq 0$ except for the trivial case. 
Thereafter, the position shift $S$ is described as the move due to the coin state; 
\begin{gather}
S \ket{x}\ket{L} := \ket{x-1}\ket{L} \\ 
\ S \ket{x}\ket{R} := \ket{x+1}\ket{R}. 
\end{gather}
Therefore, the unitary operator describing the one-step time evolution for the QW is defined as $U = S (I \otimes H)$. 
We repeat this procedure keeping the quantum coherence between the position and coin states. Finally, we obtain the probability distribution 
on the position $x$ at $t$ step as 
\begin{equation}
\Pr (X_t = x) = \Tr \left[ \left( \Tr_{c} U^{t} \left( \kb{0} \otimes \rho_0 \right) U^{t \dagger} \right) \kb{x} \right],
\end{equation} 
where $X_t$ means a random variable at $t$ step since the measurement outcome of the position measurement is probabilistically determined due to the Born rule. Here,  
$\Tr_c$ expresses the partial trace for the coin state. 

Physically speaking, the DTQW can express the free Dirac equation~\cite{Bracken, Strauch,Sato} as follows. 
We assume the specific quantum coin flip:
\begin{equation}
	H (\epsilon) = \left( \begin{array}{cc} \cos \epsilon & - i \sin \epsilon \\ - i \sin \epsilon & \cos \epsilon \end{array} \right),
\end{equation}
where the tiny parameter $\epsilon$ expresses one lattice size. Let $\Psi_n (x) = [\psi_{n,L} (x), \psi_{n,R} (x) ]^{T}$ be a coin state associated with a 
position $x$ at $n$ step. More explicitly, the coin state associated with the position $x$ at $n$ step is given by 
\begin{align}
	& U^{t} \left( \kb{0} \otimes \rho_0 \right) U^{t \dagger} \notag \\
	& = \sum_{x} \left[ \psi_{n,L} (x) \kb{x, L} + \psi_{n,R} (x) \kb{x,R} \right].
\end{align} 
The dynamics of the DTQW is given by  
\begin{align}
\Psi_n (x) & = Q (\epsilon) \Psi_{n-1} (x - \epsilon) + P (\epsilon) \Psi_{n-1} (x + \epsilon) \notag \\ 
& \approx Q(\epsilon) \left( 1 - \epsilon \frac{\partial}{\partial x} \right) \Psi_{n-1} (x) \notag \\ 
& \ \ \ \ \ \ \ \ \ + P(\epsilon) \left(1 + \epsilon \frac{\partial}{\partial x} \right) \Psi_{n-1} (x)
\end{align}
where 
\begin{gather}
Q (\epsilon) \approx \left( \begin{array}{cc} 0 & 0 \\ 0 & 1 \end{array} \right) - i \epsilon \left( \begin{array}{cc} 0 & 0 \\ 1 & 0 \end{array} \right), \\ 
P (\epsilon) \approx \left( \begin{array}{cc} 1 & 0 \\ 0 & 0 \end{array} \right) - i \epsilon \left( \begin{array}{cc} 0 & 1 \\ 0 & 0 \end{array} \right).
\end{gather} 
Therefore, we obtain the dynamics of the DTQW from the beginning, 
\begin{equation}
\Psi_\tau (x) \approx e^{- i \left(\sigma_x + \sigma_z \frac{\partial}{\partial x} \right) t} \Psi_{0} (x), 
\end{equation}
where $\sigma_x$ and $\sigma_z$ are the Pauli $x$ and $z$ matrices and $t = \epsilon \tau$. 
This equation corresponds to the Dirac equation by taking the coin state as the spinor.
Also, by developing quantum technologies to build up the precise 
measurement techniques and highly control the quantum system, it is possible to experimentally realize the DTQWs such as the 
{\it trans}-crotonic acid using the nuclear magnetic resonance~\cite{Ryan}, 
the Cs atoms trapped in the optical lattice~\cite{Karski}, the photons by employing the fiber network loop~\cite{Photon}, 
and the $^{40}{\rm Ca}^{+}$ atoms in the ion trap~\cite{Roos}.
\section{Discrete Time Quantum Walk with Periodic Position Measurement} \label{ours}
\begin{figure*}[t]
\centering
\includegraphics[width=13cm]{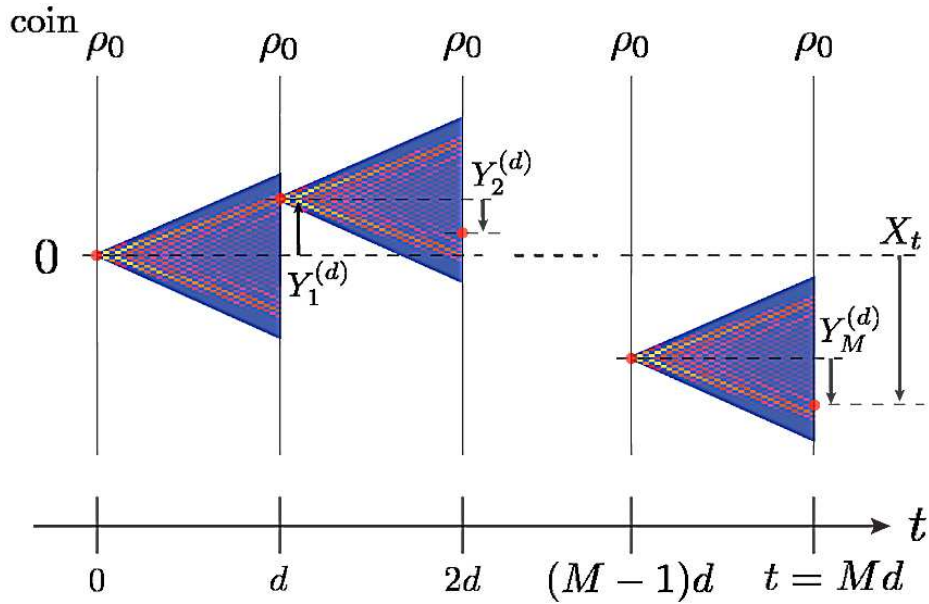}
\caption{(Color online) Quantum walk with the periodic position measurement: The schematic figure of the DTQW with PPMs is explained in the main text.}
\label{fig:1}
\end{figure*}
\begin{figure}[ht]
\centering
\includegraphics[width=7cm]{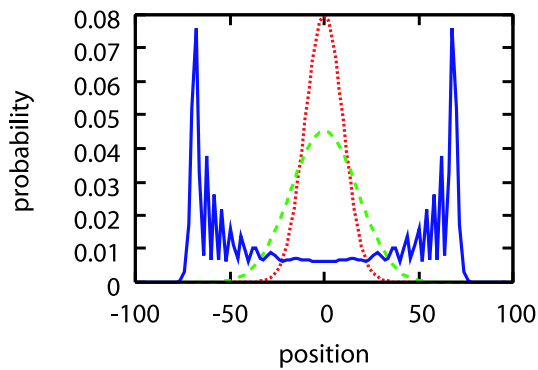}
\caption{(Color online) Example of the probability distributions at the $100$th step: The case of $\beta = 0$ corresponds to the RW (dotted line),  
$\beta =1$ corresponds to the DTQW (solid line), and $\beta = 0.5$ corresponds to the quantum-classical crossover (dash line). 
The coin parameters are $a = b = c = -d = 1/ \sqrt{2}$, which is called the Hadamard (quantum) walk. 
In the case of $\beta = 0.5$, the distribution is the Gaussian but the variance is larger than that of the RW, $\beta = 0$.}
\label{fig:2}
\end{figure}

In this section, we will show the relationship between the discrete time quantum walk and decoherence and give the interpretation to quantum foundations.
From now on, we define a simple model of the QW with environment as illustrated in Fig. \ref{fig:1}. 
After $d$ step of the DTQW, we only measure the position of the particle, that is, we take 
the projective measurement on the position state after taking the partial trace on the coin state. 
For simplicity, after the position measurement, we re-prepare the initial coin state $\rho_0$. We repeat this procedure $M$ times.
The probability distribution at the final time $t = d M$ is concerned in the following. 
This model is called the DTQW with the PPM~\cite{comment2}. We denote the sequence of the random variables on the DTQW 
between measurements by $d$ step as $\{ Y^{(d)}_i \}$. The position measurement corresponds to 
the collision with another particle, that is, environment, as in the classical sense. Furthermore, since the initial coin states are the same 
for each block of the DTQW, $\{ Y^{(d)}_i \}$ is an independent identically distributed (i.i.d.) sequence. 
The random variable for the QW with the PPM 
is denoted as $X_t = Y^{(d)}_1 + Y^{(d)}_2 + \cdots + Y^{(d)}_M$. 
When the measurement step $d$ is independent of the final time $t$, the sequence of the blocks of the DTQW by $d$ step 
can be taken as the Markov process since $\{ Y^{(d)}_{i} \}$ is an i.i.d. sequence. This case can correspond to the RW. 
To extract more detailed information, we assume that the measurement step $d$ depends on the final time $t$ as 
$d \sim t^{\beta}$ with $\beta \in [0,1]$, which is called a time scale transformation. Because of $M \sim t^{1-\beta}$, 
this means that the number of the position measurements is changed by the final time $t$. For example, 
in the case of $\beta = 0.5$, $t = 100$ step, and $d = t^{\beta}$, we take the position measurement $10$ times by $10$ steps 
and then obtain the probability distribution shown in Fig. \ref{fig:2}. 
In the following, we denote the random variable $X_t$ as $X^{(\beta)}_t$. Then, we present the following theorem 
on the limit distribution for any $\beta \in [0,1]$. 
\begin{thm} \label{theorem}
Let $\{ Y^{(d)}_i \}$ be an i.i.d. sequence of the DTQW on $\mathbb{Z}$ with the initial localized position 
$x = 0$, the initial coin qubit $\rho_0 = ( \kb{L} + \kb{R})/2$, and the quantum coin flip 
$H=a|L\rangle\langle L|+ b|L\rangle\langle R|+c|R\rangle\langle L|+d|R\rangle\langle R| \in U(2)$ 
noting that $abcd \neq 0$. Let $X_t = \sum^{M}_{i=1} Y^{(d)}_i$ be a random variable on a position 
with $d$ step between measurements and the number of the measurements $M$ with the final time $t=dM$. 
If $d \sim t^{\beta}$, then, as $t \to \infty$, we have the limit distribution as follows:
\begin{equation}
\frac{X^{(\beta)}_t}{t^{(1 + \beta)/2}} 
\Rightarrow
\left \{
\begin{array}{ll}
N(0, 1) & {\rm for} \ \beta = 0, \\
N(0, 1 - \sqrt{1 - |a|^2}) & {\rm for} \ 0 < \beta < 1, \\
K(|a|) & {\rm for} \ \beta = 1,
\end{array}
\right.
\label{main_result}
\end{equation}
where ``$\Rightarrow$" means the convergence in distribution and $N(m,\sigma^2)$ expresses the normal distribution with
the mean $m$, the variance $\sigma^2$. Note that, the random variable $K(r)$ has the probability density function $f (x ; r)$ with a parameter $r \in (0,1)$: 
\begin{equation}
	f (x ; r) = \frac{\sqrt{1-r^2}}{\pi(1-x^2)\sqrt{r^2-x^2}}\, I_{(-r,r)}(x) ,
\end{equation} 
where $I_{(-r,r)} (x)$ is the indicator function, that is, $I_A (x) = 1 \ (x \in A), =0 \ (x \notin A)$. 
\end{thm}
\begin{figure}[ht]
\centering
\includegraphics[width=4.5cm]{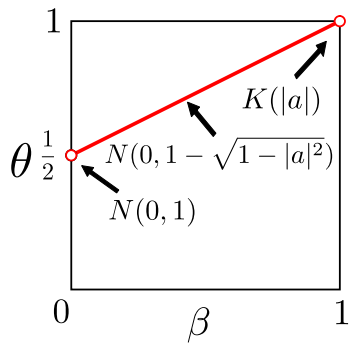}
\caption{(Color online) Limit distributions of the quantum walk with the periodic position measurement under the time scale transformation: 
The relationship between the order of the convergence time $\theta$ and the measurement time period $\beta$ in Eq. (\ref{main_result}) is illustrated.
Here, this relationship is give by $\theta = (1+\beta)/2$.}
\label{fig:3}
\end{figure}
The proof is seen in Appendix~\ref{appen}. This theorem is our main result and is illustrated in Fig. \ref{fig:3}. The theorem tells us that 
the QW with the position measurement always has the normal limit distribution. It should be noted that the case of $\beta = 1$ is the DTQW without position measurement.
The position measurement produces randomness in the QW such as the Brownian motion. 
Therefore, we always obtain that the limit distribution is the normal distribution since we always face the 
decoherence effect due to environment in the realistic experimental setup. It is extremely difficult 
to experimentally show the properties of the QW as some physical process since we only obtain the distribution after 
many steps due to $1$ step $\ll 10^{-21}$ sec in the case of the relativistic electron~\cite{Bracken}. However, the projective measurement does not uniformly 
produce randomness from the case of $0 < \beta < 1$. In other words, the parameter $\beta$ may be an indicator 
of decoherence in QWs. It is possible to evaluate the degree of decoherence from the behavior of 
the variance in the experimentally realized QW~\cite{Ryan,Karski,Photon,Roos}.

Also, the theorem tells us that the projective measurement does not always lead quantum systems to 
be classical. According to the Copenhagen doctrine, the properties of the projective 
measurement are to forget a measurement outcome and cause the quantum-classical transition, which is often called the ``state reduction". 
Therefore, the description about the quantum dynamics with many projective measurements seems to be always taken as a Markov process~\cite{Open}. 
However, from the difference of the time scale order, 
the limit distribution does not correspond to that of a Markov process like the Brownian motion with the time scale transformation although we undertake the 
projective measurements in the case of $0 < \beta < 1$. From the above fact, we give a new insight to the arrow of time. Except for $\beta = 1$, we 
undertake the position measurement for the DTQW infinite times. In other words, our obtained entropy by position measurement is always infinity 
as $t \to \infty$ but its increasing rate depends on the time-scale parameter $\beta$. Therefore, our result (\ref{main_result}) shows that 
the projective measurement does not uniformly produce the arrow of time for any time scale. 
Note that, this always leads to the arrow of time according to the Copenhagen doctrine~\cite{comment3}. 
\section{Conclusion} \label{conc}
We have analytically obtained the limit distribution of the DTQW 
on $\mathbb{Z}$ with PPMs (\ref{main_result}). From this 
mathematical result, we have shown that the origin of randomness is the position measurement like 
the Brownian motion but there does not always exist the quantum-classical transition in the DTQW by the 
position measurement since the degree of randomness is not time scale invariant. 
Also, we have constructed the quantification of the DTQW with PPMs. 
Furthermore, we have applied the concept of the QW to the problem of quantum measurement 
and shown that the quantum dynamics with the projective measurements cannot be taken as a 
Markov process under some specified time scales, that is, this is not time scale invariant.  
Our results show that randomness and the arrow of time in QWs gradually emerge. 
Furthermore, we will prove the limit distribution in the cases of the general 
coin state and the continuous time QW~\cite{CKSS}. 
\section*{Acknowledgment}
	The authors thank Jun Kodama and Satoshi Kawahata for useful comments and 
	one of the authors (YS) also thanks Lorenzo Maccone, Akio Hosoya, Jeffrey Goldstone, Hisao Hayakawa, Masanori Ohya, and Ra\'{u}l 
	Garc\'{i}a-Patr\'{o}n for useful discussions and comments. YS is supported by 
	JSPS Research Fellowships for Young Scientists (Grant No. 21008624).
\appendix
\section{Proof of Theorem~\ref{theorem}} \label{appen}
In this appendix, we give the proof of the theorem basically using the spatial Fourier transform~\cite{Segawa, Grimmett} and the Taylor expansions.
\begin{proof}
Since the case of $\beta = 0$ can be taken as the RW, we obtain the normal distribution as the limit distribution from 
the well-known central limit theorem. Furthermore, since the case of $\beta = 1$ corresponds to 
the DTQW, its limit distribution has already shown in Refs.~\cite{KonnoQL0,KonnoQL1}. In the following, 
we only consider the limit distribution in the case of $0 < \beta < 1$. 

Since $\{ Y^{(d)}_{i} \}$ is the i.i.d. sequence and $X_t = \sum_{i=1}^{M} Y^{(d)}_{i}$, we can describe  
$E(e^{i\xi X_t}) = \{ E(e^{i\xi Y_1^{(d)}}) \}^M$, where $E (Z)$ is the expectation value of the random variable $Z$ and $E(e^{i \xi Z})$ 
is called the characteristic function of $Z$, as
\begin{align} 
& E(e^{i \xi X_t}) \notag \\
& \ \ \ =\left\{ \int_0^{2\pi} \frac{1}{2}\mathrm{Tr}[\widehat{H}^{d}(k+\xi)\cdot\widehat{H}^{-d}(k)] \frac{dk}{2\pi} \right\}^{M}, \label{chara}
\end{align}
where $\widehat{H}(k)=(e^{ik}|R\rangle \langle R|+e^{-ik}|L\rangle \langle L|)H$, which is the spatial Fourier transform. 
Note that, the details of this method is seen in Ref.~\cite{Segawa} used in the different context. 
Let the eigenvalue of $\widehat{H}(k)$ be $e^{i\varphi_l (k)} \ (l \in \{ \pm \})$. 
Then, we can rewrite the integrand of Eq. (\ref{chara}) for replacing $\xi$ with $\xi/t^\theta$ and 
$d$ with $t^{\beta}$ in the following: when $\theta > \beta$, we have 
\begin{align} 
& \frac{1}{2}\mathrm{Tr}[\widehat{H}^{t^\beta}(k+\xi/t^\theta)\cdot\widehat{H}^{-t^\beta}(k)] \notag \\
& \approx \frac{1}{2} \sum_{l,m \in \{ \pm \} } e^{it^\beta [ \varphi_l(k+\xi/t^\theta)-\varphi_m(k) ]} \times \delta_{l,m} \notag \\
& \approx \frac{1}{2} \sum_{l,m \in \{ \pm \} } e^{it^\beta [ \varphi_l(k)-\varphi_m(k) ]} \times e^{i\xi t^{\beta-\theta} \frac{d \varphi_l (k)}{dk}} 
	\times \delta_{l,m} \notag \\
& =  \frac{1}{2} \left( e^{i \xi t^{\beta-\theta} h(k)} + e^{- i \xi t^{\beta-\theta} h(k)} \right) \notag \\
& \approx 1-\frac{\xi^2}{2t^{2(\theta-\beta)}}[h(k)]^2, \label{nakami}
\end{align}
using the Taylor expansion for sufficiently large $t$. Here, $\delta_{l,m}$ is the Kronecker delta 
and $h(k) \equiv \frac{d\varphi_{+}(k)}{dk} = - \frac{d\varphi_{-}(k)}{dk}$.
By Eqs. (\ref{chara}) and (\ref{nakami}), only if $2(\theta-\beta)=1 - \beta$, then there exists the limit distribution as follows: 
\begin{align}
E (e^{i\xi X_t^{(\beta)}/t^\theta}) 
& \approx \left\{ \int_{0}^{2\pi} \left( 1-\frac{\xi^2}{2t^{2(\theta-\beta)}}[h(k)]^2 \right)
	\frac{dk}{2\pi}\right\}^{t^{1-\beta}} \notag \\	
&= \left[ 1 - \frac{\xi^2}{2t^{2(\theta-\beta)}} \sigma (a)^2 \right]^{t^{1-\beta}} \notag \\
& \to e^{- \xi^2\sigma (a)^2 /2} \ \mathrm{as} \ t \to \infty, 
\end{align}
where 
\begin{equation}
\sigma (a)^2 \equiv \int_0^{2\pi} [h(k)]^2 \frac{dk}{2\pi}.
\label{deri}
\end{equation}
In the following, we derive $\sigma (a)^2$. 
Generally, a quantum coin flip $H \in U(2)$ is expressed by four parameters $r,\;\phi,\;\psi, \;\delta$ with $r\in (0,1)$, $\phi,\;\psi, \;\delta\in \mathbb{R}$ such that 
\begin{equation} 
H(r,\phi,\psi,\delta)= \left( \begin{array}{cc} re^{i\phi} & \sqrt{1-r^2}e^{i\psi} \\ -\sqrt{1-r^2}e^{-i(\psi-\delta)} & re^{-i(\phi-\delta)} \end{array} \right).
\end{equation}
The eigenvalues of $\widehat{H}(k)$ is given by the solution for 
\begin{equation}
	\mathrm{det}[H(r,\phi+k,\psi+k,\delta)-zI]=0.
\end{equation}
So, we have 
\begin{align}\label{solution}
& \mathrm{det}[H(r,\phi+k,\psi+k,\delta)-zI] \notag \\
&= [re^{i(\phi+k)}-z][re^{-i(\phi+k-\delta)}-z]+(1-r^2)e^{i\delta} \notag \\
&= z^2-2re^{i\delta/2} \cos (\delta'+k) z + e^{i\delta} = 0,
\end{align}
where $\delta' \equiv \phi-\delta/2$. Putting the solutions, $e^{i \varphi_{+}(k)}$ and $e^{i \varphi_{-}(k)}$, for Eq. (\ref{solution}), 
we obtain 
\begin{align}
 e^{i \varphi_{+}(k)}+e^{i \varphi_{-}(k)} &= 2re^{i\delta/2}\cos (\delta'+k) \label{kai} \\
 \varphi_{+}(k) + \varphi_{-}(k) &= \delta+2m\pi \label{kai2},
\end{align}
where $m\in \mathbb{Z}$. By Eqs. (\ref{kai}) and (\ref{kai2}), we have 
\begin{equation}\label{cos}
\cos [ \varphi_{+}(k) -\delta/2 ] =r\cos (\delta'+k). 
\end{equation}
By differentiating both sides of Eq. (\ref{cos}) with respect to $k$, 
\begin{equation}
\left[ \frac{d \varphi_{+}(k)}{dk}\right]^2 = [h(k)]^2 = \frac{r^2 \sin ^2(\delta'+k)}{1-r^2 \cos ^2(\delta'+k)}. 
\end{equation}
Therefore, we derive Eq. (\ref{deri}) as 
\begin{align}
\sigma (a)^2 &= \int_0^{2\pi} \left[ h(k) \right]^2 \frac{dk}{2\pi}=\int_0^{2\pi} 
	\frac{|a|^2\sin^2 k}{1-|a|^2\cos^2 k}\frac{dk}{2\pi} \notag \\
	&= 1-\sqrt{1-|a|^2}. 
\end{align}
This shows that $\sigma (a)^{2}$ only depends on the parameter $a$ of the quantum coin flip.
Therefore, the proof is completed.
\end{proof} 
%%%%%%%%%%%%%


\begin{thebibliography}{99}
\bibitem{comment3}
While we assume the ``Copenhagen interpretation" throughout this paper, 
there are different problems on the arrow of time in another interpretation. For example, Maccone~\cite{Lorenzo} recently showed the resolution of the arrow of time 
in the Everett relative state interpretation noted in Ref.~\cite{Everett}.
\bibitem{comment2}
Our model is different from the ``decoherence QW" analyzed by Zhang~\cite{Zhang}. 
In the decoherence QW, one often take the joint measurement of the position and the coin. 
The limit distribution of the case of $\beta = 0$, that is, the measurement period is independent of the final time, is the normal 
distribution with the convergence time order $1/2$. There remains the open problem to similarly analyze the extended Zhang model.
\end{thebibliography}
\end{document}